\documentclass[preprint2]{aastex}

\usepackage{graphicx}
\usepackage{amsmath}
\usepackage{natbib}
\usepackage{mathtools}

\usepackage{txfonts}
\providecommand{\abs}[1]{\left|#1\right|}
\newcommand{\sdoaia}{{\it SDO}/AIA\ }

\newcommand{\ipi}{\text{i}\pi}
\newcommand{\muhz}{$\mu$Hz}

\slugcomment{}

\shorttitle{}
\shortauthors{Auch\`ere et al.}

\begin{document}

   \title{THERMAL NON-EQUILIBRIUM REVEALED BY PERIODIC PULSES OF RANDOM AMPLITUDES IN SOLAR CORONAL LOOPS}

	\author{F. Auch\`ere\altaffilmark{1}, C. Froment\altaffilmark{1}, K. Bocchialini\altaffilmark{1}, E. Buchlin\altaffilmark{1} and J. Solomon\altaffilmark{1}}
\affil{Institut d'Astrophysique Spatiale, CNRS, Univ. Paris-Sud, Universit\'e Paris-Saclay, B\^at. 121, 91405 Orsay, France}
   \email{frederic.auchere@ias.u-psud.fr}

\date{\it \center{Received 2016 April 14; revised 2016 June 7; accepted 2016 June 16}}

\begin{abstract}
We recently detected variations  in extreme ultraviolet intensity in coronal loops repeating with periods of several hours. Models of loops including stratified and quasi-steady heating predict the development of a state of thermal non-equilibrium (TNE): cycles of evaporative upflows at the footpoints followed by falling condensations at the apex. Based on Fourier and wavelet analysis, we demonstrate that the observed periodic signals are indeed not signatures of vibrational modes. Instead, superimposed on the power law expected from the stochastic background emission, the power spectra of the time series exhibit the discrete harmonics and continua expected from periodic trains of pulses of random amplitudes. These characteristics reinforce our earlier interpretation of these pulsations as being aborted TNE cycles.
\end{abstract}
  
%

\section{THE THERMAL NON-EQUILIBRIUM (TNE) DEBATE\label{sec:_tne_debate}}


The very existence of solar and stellar coronae remains one of the great problems in astrophysics. In particular, the heating mechanism(s) capable of keeping the plasma confined in magnetic loops at temperatures of several million degrees still resist comprehensive understanding \citep{Klimchuk2015}. Despite considerable effort and progress \citep[for a review, see][]{Reale2014}, we do not know for sure where the heating occurs or how it evolves with time. In the Parker field-line tangling scenario \citep{Parker1972, Parker1988}, one can expect the heating to be highly stratified, i.e. concentrated at the footpoints of the loops \citep{Rappazzo2007}. If in addition it is quasi-steady, i.e. varying slowly (or impulsively with a high repetition rate) compared to the cooling time,  numerical simulations consistently show that the loops are susceptible to entering a regime of thermal non-equilibrium \citep[e.g.][]{Kuin1982, Antiochos1991, Karpen2001, Muller2003, Mok2008, Klimchuk2010, Lionello2013}.
For specific combinations of the heating conditions and geometry, the footpoint heating drives evaporative upflows, hot plasma accumulates in the loop, and as it cools, a condensation grows quickly near the apex, falls down one leg, hits the chromosphere, and the cycle repeats with periods from several tens of minutes to several hours. 

This process is thought to play a significant role in the formation of prominences \citep{Antiochos1991, Karpen2006} and coronal rain \citep{Muller2003, Muller2004, Muller2005, Antolin2010, Antolin2015}. However, 
 \cite{Klimchuk2010} argued that TNE models fail to reproduce simultaneously the key observational properties of coronal loops, thus discarding the possibility that highly stratified, quasi-constant heating could be the norm in active regions. But other studies \citep{Mikic2013, Lionello2013, Lionello2016, Winebarger2014} have shown that the inconsistencies with observations can be resolved if the geometry is more complex than the constant cross-section, semicircular vertical loops used by \cite{Klimchuk2010}. In particular, if the loops are expanding and asymmetric, the condensations do not fully develop. The plasma thus remains at coronal temperatures and densities, which results in unstructured intensity profiles, as observed in the extreme ultraviolet (EUV). 
Still, this does not prove that quasi-steady stratified heating is commonplace, because outside TNE conditions it leads to hydrostatic solutions that, at least for monolithic loops, seem incompatible with several observational constraints~\citep{Reale2014}. Other scenarios involving more sporadic heating, such as the nanoflare storms,  	have been developed to resolve these issues \citep{Klimchuk2009, Viall2013}.

Surprisingly, in this debate, the most striking characteristic of TNE conditions -- the predicted periodicity of the temperature and density and hence of the plasma emissivity -- has not yet been searched for in the observations in order to test the models.
We processed more than 13 years of observations at 19.5~nm with the Extreme-ultraviolet Imaging Telescope \citep[EIT;][]{Delaboudiniere1995} of the {\it Solar and Heliospheric Observatory} \citep[{\it SOHO};][]{Domingo1995} and discovered hundreds of long-period (3-16 hr) pulsation events in coronal loops, some lasting for up to six days \citep{Auchere2014}. \citet{Froment2015} analyzed in detail three other events observed in the six coronal bands of the Atmospheric Imaging Assembly \citep[AIA;][]{Lemen2012} on the {\it Solar Dynamics Observatory} \citep[{\it SDO};][]{Pesnell2012}. The differential emission measure (DEM) tools developed by \citet{Guennou2012a, Guennou2012b, Guennou2013} revealed periodic variations of the total emission measure and DEM peak temperature that resemble those in the TNE simulations of \citet{Mikic2013}. 


However, despite the similarities, we cannot yet unambiguously conclude that the observed pulsations are caused by TNE without additional evidence, such as spectroscopic observations of the predicted outflows. For example, while no magnetohydrodynamic mode can explain periods of several hours in coronal loops \citep{Auchere2014}, it is difficult to exclude the possibility of slow beats resulting from the coupling between adjacent loops of similar eigenfrequencies. But in this paper, we demonstrate that the power spectral densities (PSD) of the time series in which \citet{Froment2015} detected pulsations do not have the characteristics expected from waves or damped waves, but instead those of signals known as random pulse trains. This reinforces the idea that the observed system undergoes a cyclic evolution in a constantly varying environment, as expected if TNE is at play in coronal loops.



\section{WAVES {\it VS.} PULSE TRAINS\label{sec:waves_vs_pulses}}

\begin{figure*}
	\centering
		\includegraphics[width=\textwidth]{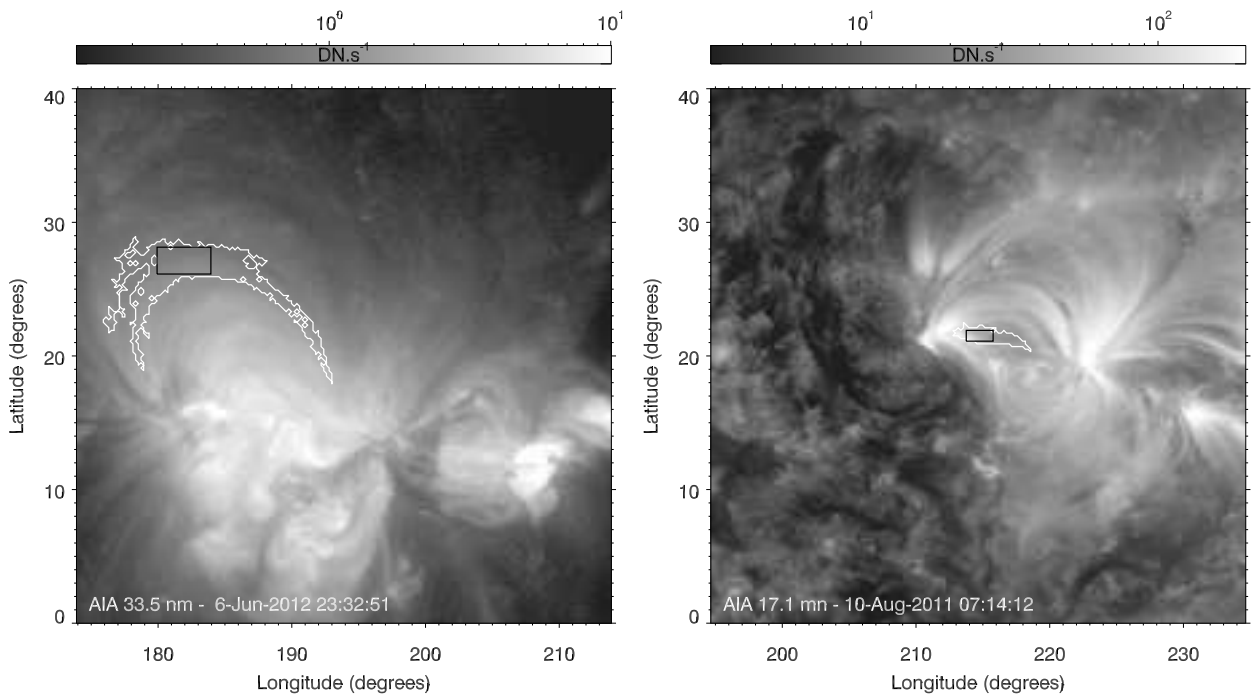}
		\caption{Left: middle frame of the one-minute cadence, 6.4 day-long sequence corresponding to Case 1 of \citet{Froment2015}. Right:  middle frame of the one-minute cadence, 4 day-long, AIA 17.1~nm sequence corresponding to Case 3 of \citet{Froment2015}. The regions where excess Fourier power was automatically detected (white contours) delineate two bundles of loops, one in the outskirts of NOAA AR 11499 (left), the other one in the core of NOAA AR 11268 (right). Figures~\ref{fig:case_1_335} and~\ref{fig:case_3_171} present the time series obtained by averaging the intensity over the black boxes, along with their Fourier and wavelet power spectra.}
	\label{fig:aia_mid_frames}
\end{figure*}

The Fourier power spectra of  time series of coronal intensity commonly exhibit an overall power-law behavior caused by a background of stochastic plasma processes \citep{Gruber2011, Auchere2014, Inglis2015}. A hump superimposed on this basic shape is also frequently observed \citep{Ireland2015, Auchere2016}. Two examples of such power spectra are given in the rightmost panels of Figures~\ref{fig:case_1_335} and~\ref{fig:case_3_171} (gray histograms), the corresponding time series being shown in the top left panels. These latter have been obtained from two sequences of \sdoaia images (Figure~\ref{fig:aia_mid_frames}) by averaging the  intensity over the regions selected by \citet{Froment2015} for plasma diagnostics (black boxes). These time series and their spectral analysis are described in detail in \S~\ref{sec:aia_detection}.

There are two fundamentally different possibilities to explain the humps in the power spectra. First, they can be due to periodic damped oscillations: from the convolution theorem, the power spectrum of a damped wave is obtained from the convolution of the Fourier transform of the damping function with that of the wave. For example, the power spectrum of an exponentially decaying sine is a Lorentzian centered on the sine frequency. While the resulting hump represents excess power compared to a background power law, its presence is not sufficient in itself to infer the presence of a periodic phenomenon. Indeed, the second possibility is that the hump is due to the presence in the time series of one or a few pulses\footnote{The term {\it pulse} is used throughout the paper to describe a rapid, transient increase in intensity followed by a rapid return to the original value.} of similar widths, even if they are not periodic, as in the reference region of \citet{Auchere2016}. For example, the power spectrum of a single exponential pulse is also a Lorentzian, but unlike the case for a damped wave -- and this is a major difference -- the hump is now centered at zero frequency. 

In all the cases that we have studied \citep[and also in the moss regions examined by][]{Ireland2015}, the width of the hump is comparable to its central frequency. For example, fitting the power spectrum of the rightmost panel of Figure~\ref{fig:case_1_335} with the sum of a power law and a Gaussian without forcing the latter to be centered at zero, yields a central frequency of 26~\muhz\ and a full width at $1/e$ of 31~\muhz. If interpreted as a damped wave, this would correspond to a damping time shorter than the period itself. In addition, as shown in \S~\ref{sec:aia_detection}, a Gaussian centered at zero is justas valid a fit of this PSD. Therefore, the best explanation for the presence of several peaks (more than 15 in the time series of Figures~\ref{fig:case_1_335} and~\ref{fig:case_3_171}) is that the physical phenomenon at their origin repeats itself, potentially with different initial and boundary conditions each time. This leads to the idea that the time series should in fact be interpreted as a periodic succession of pulses of random amplitudes. 


\section{PSDs OF RANDOM-AMPLITUDE PULSE TRAINS\label{sec:pulse_trains_psds}}

\begin{figure}[t]
	\centering
		\includegraphics[width=0.456\textwidth]{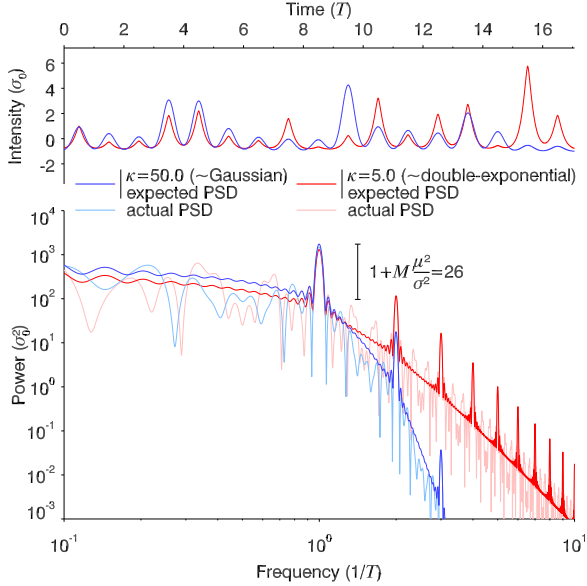}
		\caption{Top: sample random pulse trains for rounded ($\kappa=50$, blue) and pointed ($\kappa=5$, red) pulses defined by Equation~\ref{eq:kappa_pulse}. Bottom: corresponding expected (red and blue) and actual (lighter shades of red and blue) PSDs. The $\chi_3^2$ distribution of amplitudes creates a continuum that is a scaled version of the PSD of the elementary pulse. The contrast between the spectral lines and the continuum  depends only on the number of pulses and on the statistical distribution of their amplitudes.}
	\label{fig:sim_pulse_train}
\end{figure}

A periodic succession of pulses of random amplitudes, called a random pulse train, can be expressed as
\begin{equation}
f(t) = \sum_{m=0}^{M-1} a_m p(t - mT),
\label{eq:pulse_train_from_annex}
\end{equation}
\noindent
where $t$ is the time and $T$ is the repetition period of $M$ copies of an elementary pulse $p(t)$ with random amplitudes $a_m$.
The corresponding expected PSD is given by (see the Appendix and references therein)
\begin{equation}
\Psi(\nu) = \abs{P(\nu)}^2\left\lbrace M\sigma^2 + \mu^2\left(\frac{\sin(\pi\nu TM)}{\sin(\pi\nu T)}\right)^2\right\rbrace
\label{eq:random_pulse_train_power_from_annex}
\end{equation}
\noindent
where $\abs{P(\nu)}^2$ is the power spectrum of the elementary pulse $p(t)$, and $\sigma$ and $\mu$ are respectively the standard deviation and the mean of the statistical distribution of the amplitudes of the pulses. The power spectrum of the pulse train is thus the power spectrum of the elementary pulse modulated by a function periodically peaked in frequency with period $1/T$.
\begin{figure*}
	\centering
		\includegraphics[width=\textwidth]{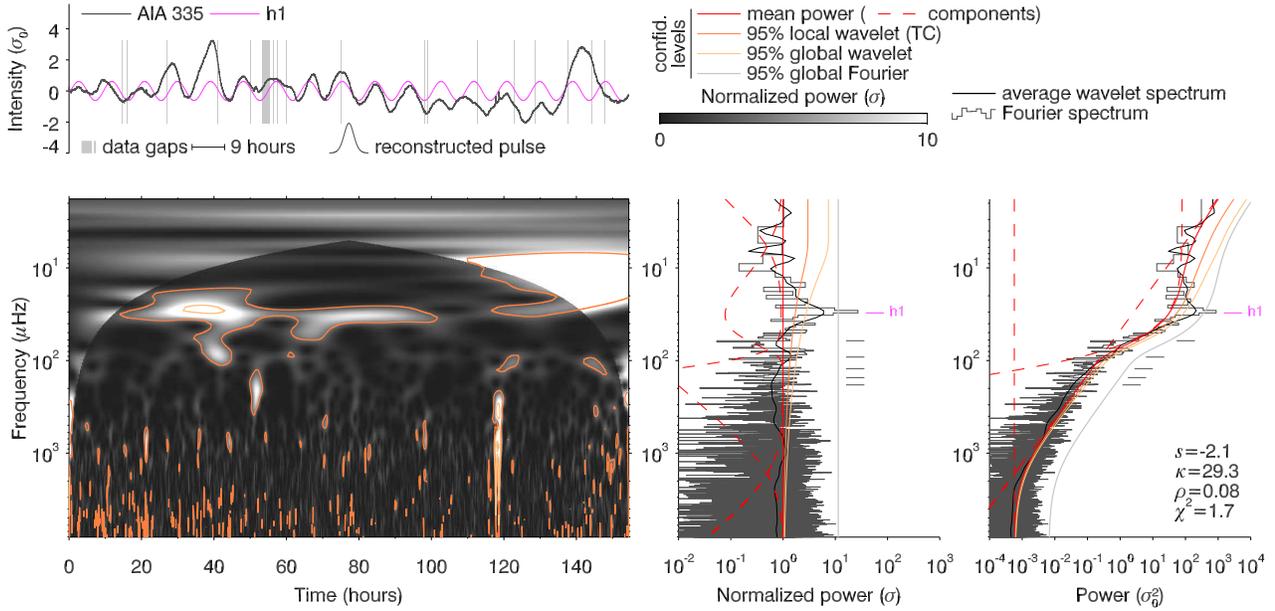}
		\caption{The time series corresponding to the left panel of Figure~\ref{fig:aia_mid_frames} is shown in the top left panel. Its Fourier and time-averaged wavelet power spectra (rightmost panel, gray histograms and black curves) exhibit a broad hump superimposed on a power law leveling off at high frequencies. The 26.3~$\sigma$ peak of Fourier power labeled h1 at 30~\muhz\ stands out in the whitened spectra (middle panel) and has a probability of random occurrence of $1.7\times 10^{-8}$. The corresponding Fourier component is overplotted on the time series in magenta. The whitened wavelet spectrum (left panel) shows a matching strip of significant power lasting for most of the sequence. The elementary pulse reconstructed by inverse Fourier transform of the  kappa function component (dashed red) of the mean power fit (solid red) resembles the shape of the pulsations in the light curve. Power within the cone of influence of the Morlet wavelet is shown in lighter shades of gray.}
	\label{fig:case_1_335}
\end{figure*}

\begin{figure*}
	\centering
		\includegraphics[width=\textwidth]{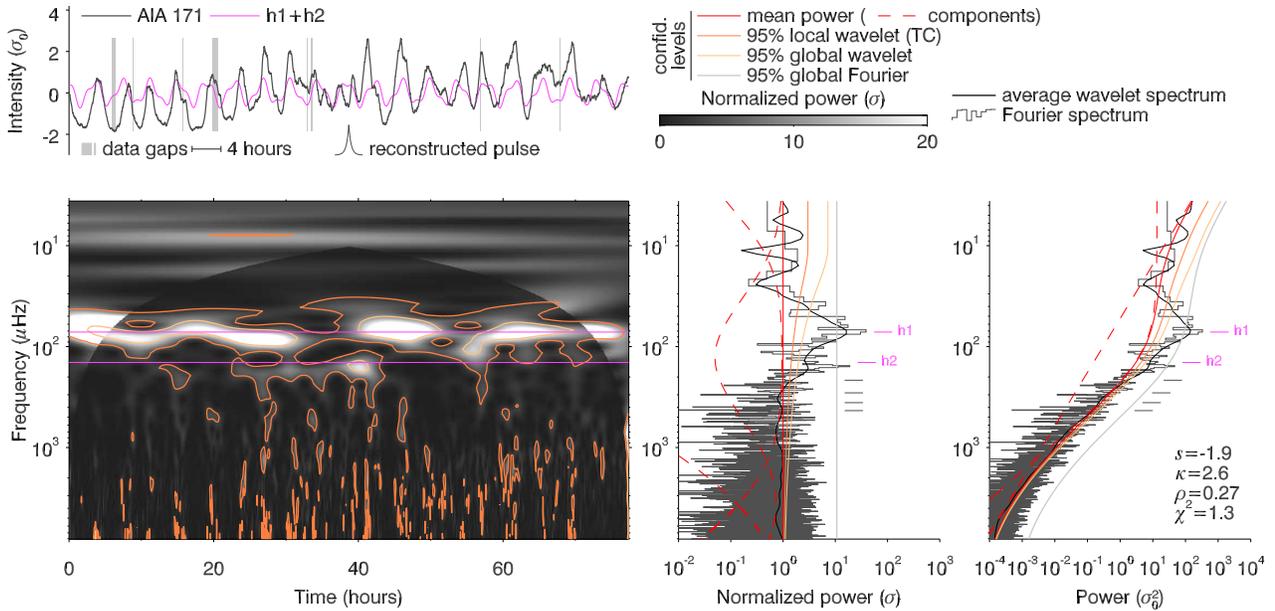}
		\caption{Same as Figure~\ref{fig:case_1_335} but for the time series corresponding to the right panel of Figure~\ref{fig:aia_mid_frames}. }
	\label{fig:case_3_171}
\end{figure*}

\cite{Auchere2016} found that the PSDs of many coronal time series can be represented by the following model: 
\begin{equation}
\sigma(\nu)=A\nu^s+B\text{K}_{\rho,\kappa}(\nu)+C,
\label{eq:kappa_model}
\end{equation}
\noindent
where the first term is a power law of slope $s$ representing the background power, the second term is a kappa function representing the hump, and the third term is a constant representing high-frequency white noise. In order to illustrate the properties of the power spectra given by Equation~\ref{eq:random_pulse_train_power_from_annex}, we thus consider trains whose elementary pulses $p(t)$ have power spectra proportional to the kappa function term of the above power model,
\begin{equation}
\abs{P(\nu)}^2=\text{K}_{\rho,\kappa}(\nu)=\left(1+\frac{\nu^2}{\kappa \rho^2}\right)^{-\frac{\kappa+1}{2}}.
\label{eq:kappa_psd}
\end{equation}
\noindent
The analytic expression of these pulses is obtained by taking the inverse Fourier transform of the square root of the kappa function:\footnote{The expression was obtained with the Mathematica software.}
%
\begin{equation}
p(t) =\frac{2\pi^{\frac{\kappa+1}{4}}(\kappa\rho^2)^{\frac{\kappa+3}{8}}}{\Gamma\left(\frac{\kappa+1}{4}\right)} \abs{t}^{\frac{\kappa-1}{4}}K_{\frac{\kappa-1}{4}}\left(2\pi\rho\sqrt{\kappa}\abs{t}\right)
\label{eq:kappa_pulse}
\end{equation}

\noindent
where $K_\alpha(x)$ denotes the modified Bessel function of the second kind and $\Gamma(x)$ denotes the gamma function. The initial fraction ensures normalization to unity. For a given width $\rho$, the pulses tend to a Gaussian as $\kappa$ tends to infinity, and they become increasingly peaked as $\kappa$ decreases. For $\kappa=3$ the pulse is a double-exponential.


%



Two sample pulse trains, normalized to their standard deviation $\sigma_0$, are plotted in the top panel of Figure~\ref{fig:sim_pulse_train} as a function of $t/T$, in blue for rounded (nearly Gaussian, $\kappa=50$) pulses and in red for pointed (nearly double-exponential, $\kappa=5$) pulses. Apart from the pulse shape, all parameters are equal: $M=17$ pulses equally spaced by $T$, of width\footnote{Corresponding to a full width at half maximum of $\approx T/2$ for $\kappa=50$ and $\approx T/3$ for $\kappa=5$.} $\rho=T/120$, and of amplitudes drawn -- as an example and to ensure positiveness -- from a chi-squared distribution of degree 3, which has mean $\mu=3$ and variance $\sigma^2=6$. The bottom panel shows, with the same color coding, the corresponding expected PSDs\footnote{Note that any distribution of amplitudes with the same coefficient of variation $\sigma/\mu$ would result in identical expected PSDs (see the Appendix and Equation~\ref{eq:random_pulse_train_power}).} computed using Equation~\ref{eq:random_pulse_train_power_from_annex} after substitution of $\abs{P(\nu)}^2$ by its expression given in Equation~\ref{eq:kappa_psd}. They are the {\it average} PSDs that one would expect from an infinite number of realizations of the amplitudes, {\it not} the PSDs of the curves of the top panel.\footnote{The analytic expressions for the PSDs of the two particular pulse trains of the top panel of Figure~\ref{fig:sim_pulse_train} can be derived from the Fourier transform of Equation~\ref{eq:kappa_pulse} and the time-shifting theorem (Equation~\ref{eq:generic_pulse_train_power}). They are represented in light shades of blue and red, but they are of no practical use.} Since the number of pulses, the period, and the distribution of amplitudes are identical in both cases, so is the periodic modulation term between brackets in Equation~\ref{eq:random_pulse_train_power_from_annex}. Only the PSDs of the elementary pulses -- which correspond to the lower envelopes -- are different. For nearly Gaussian pulses, $\abs{P(\nu)}^2$ is also nearly Gaussian, while for nearly double-exponential pulses, $\abs{P(\nu)}^2$ has an extended high-frequency power-law wing of slope -6. 

Interestingly, 
while the harmonic peaks are due to the periodicity of the pulses, the continuum $M\sigma^2\abs{P(\nu)}^2$ arises from the randomness of their amplitudes ($\sigma\neq 0$). The probability density function (PDF) of the amplitudes of an observed pulse train cannot be completely determined from the PSD because only the mean and variance  appear in Equation~\ref{eq:random_pulse_train_power_from_annex}. 
Nonetheless, for a given number of pulses, the contrast between the peaks and the continuum, given by
\begin{equation}
1 + M\frac{\mu^2}{\sigma^2},
\label{eq:random_pulse_contrast}
\end{equation}
\noindent
provides the coefficient of variation $c_v=\sigma/\mu$, which quantifies the extent of variability of a random variable in relation to its mean. While the contrast increases with the number of pulses, it decreases with the square of the coefficient of variation, which implies that the periodic component of the PSD tends to vanish if the amplitudes are highly variable. In the example of Figure~\ref{fig:sim_pulse_train}, the PDF is $\chi_3^2$, $c_v=\sqrt{2/3}$ and the contrast is $1+3M/2=26.5$.

The only signature of the periodicity of the pulses in the PSD is the presence of harmonic peaks. Since the PSD of a single pulse is proportional to $\abs{P(\nu)}^2$, it is indistinguishable from the continuum of the PSD of a random pulse train. Therefore, the hump in the PSD of a real time series should in all cases be accounted for as background power. This is essential in order to derive proper confidence levels for the detection of the peaks, and justifies {\it a posteriori} the model of Equation~\ref{eq:kappa_model}.



\section{EVIDENCE FOR PULSE TRAINS IN CORONAL LOOPS\label{sec:pulses_evidence}}
\subsection{Detection In AIA Data.\label{sec:aia_detection}}

In this section, we re-examine two of the three cases (Case 1 and Case 3) studied in detail  by~\citet{Froment2015} in the light of the properties of random pulse trains described in \S\ref{sec:pulse_trains_psds}. We picked these two cases because, as demonstrated  in \S\ref{sec:comparison_simulations}, they exhibit the two types of pulses shown in Figure~\ref{fig:sim_pulse_train}: nearly Gaussian and nearly double-exponential. Figure~\ref{fig:aia_mid_frames} shows the middle frames of the two  input AIA sequences. Case 1 corresponds to the one-minute cadence, 9202 frame-long, 33.5~nm sequence (left panel), starting 2012 June 3 at 18:00~UT and ending 2012 June 10 at 04:29~UT. Case 3 corresponds to the one-minute cadence, 4611 frame-long, AIA 17.1~nm sequence (right panel) starting 2011 August 8 at 04:01 UT and ending 2011 August 12 at 03:59~UT. Each original image has been binned over $4\times 4$ pixels and remapped in heliographic coordinates \citep{Auchere2005} with a $0^{\circ}.05$ sampling pitch in longitude and latitude for feature tracking. 
The white contours delineate the regions of excess Fourier power \citep[Figure 4 of][]{Froment2015}. The time series have been obtained by averaging the intensity over the black boxes. The Fourier and wavelet analyses of these time series, including a critical reassessment of confidence levels, have been described in detail in \citet{Auchere2016} and are summarized below.

The time series of Case 1 is plotted in dark gray in the top left panel of Figure~\ref{fig:case_1_335}. Data gaps, defined as the intervals during which no data exist within 30 s of an integer number of minutes since the beginning, represent 0.7\% of the sequence and are represented by the vertical gray bars, the height of which also represents the range of variation of the intensity. The gaps have been filled with linear interpolations between the nearest data points. Since we used a one-minute cadence sample of the original 12~s cadence AIA data, the remainder of the time series was considered to be evenly spaced and thus kept as-is. The histogram-style curve of the right panel is the Fourier power spectrum of the Hann-apodized time series. The solid red curve is the least-squares fit (of reduced $\chi^2=1.7$) of this spectrum with the three-component (dashed red curves) model $\sigma(\nu)$ of Equation~\ref{eq:kappa_model}. The hump formed by the kappa function term dominates the expected background power law between 6 and 80~\muhz. The peak of Fourier power at 30~\muhz\ (9 hr) labeled h1 exceeds the 95\% {\it global}\footnote{{\it Global} confidence levels take into account the total number of degrees of freedom in the spectra, as opposed to {\it local} confidence levels that apply to individual frequencies and/or dates \citep{Auchere2016}.} confidence level (gray curve) and reaches 26.3~$\sigma$, which corresponds to probability of random occurrence of $1.7\times 10^{-8}$ \citep{Scargle1982, Auchere2016}. The same information is displayed in the middle panel after whitening of the spectrum, i.e. normalization to $\sigma(\nu)$.

The bottom left panel shows the whitened wavelet spectrum of the zero-padded time series. The power at 30~\muhz\ (magenta line) exceeds the 95\% {\it local} confidence level (orange contours) during most of the sequence, with a maximum above the 95\% {\it global} confidence level (yellow contours) 39~hr after the beginning. Such a long-lived structure has a probability of random occurrence of $7\times 10^{-11}$. This produces a 6~$\sigma$ peak in the time-averaged wavelet spectrum (black curves in the middle and right panels) that lies above the 95\% global confidence levels (yellow curves), with an associated random occurrence probability of $6\times 10^{-7}$.

Figure~\ref{fig:case_3_171} is identical to Figure~\ref{fig:case_1_335} for Case~3. In the fitted model (red curves), the kappa function dominates the power-law between 13 and 3000~\muhz. The peak of Fourier power at 72~\muhz\ (3.9 hr) labeled h1 exceeds the 95\% global confidence level (gray curves) and reaches 38.9~$\sigma$, which corresponds to a probability of random occurrence of $1.5\times 10^{-14}$. The  power at 72~\muhz\ in the wavelet spectrum of the bottom left panel exceeds the 95\% global confidence level (yellow contour) during most of the sequence. This produces a 16.3~$\sigma$ peak in the time-averaged wavelet spectrum (black curves). The associated probabilities are too low to be meaningful.

A second peak of Fourier power surpasses the 95\% global confidence level at 158~\muhz\ (1.8 hr). At 18.9~$\sigma$, it has a probability of random occurrence  of $1.4\times 10^{-5}$. It lies 14~\muhz, or 8\%, higher than the theoretical frequency of the second-order harmonic -- labeled h2 -- of the primary peak (h1, the fundamental, or first harmonic). The expected frequencies of the higher undetected orders are marked by gray ticks. The h2 peak corresponds in the wavelet spectrum to the secondary band of power that exceeds the 95\% local confidence level between 23 and 43~hr after the beginning of the sequence, preceded by an isolated peak  at the same frequency around 15~hr. The secondary band of wavelet power actually lies at exactly twice the frequency of the fundamental between 23 and 31~hr, both peaks being shifted by about 14~\muhz\ toward the high frequencies compared to h1 and h2 (magenta lines). It is thus likely that the secondary peak of Fourier power is indeed the second harmonic, the offset from h2 resulting from a combination of the noise and of the temporal variations of the fundamental frequency. As we will see in the next section, this explanation is corroborated by the more pointed shape of the pulses at the times where the harmonic is visible in the wavelet spectrum. Other explanations would require either a physical mechanism of frequency-doubling or the presence along the line of sight of a second structure pulsating at twice the frequency of the other.


Combined with our analysis of the possible artefacts~\citep{Auchere2014}, all confidence levels indicate beyond reasonable doubt that the periodicities detected in the two time series of Figures~\ref{fig:case_1_335} and~\ref{fig:case_3_171} are of solar origin. Unlike in most observational studies of coronal loops
, we did not subtract an estimate of the background and foreground emission. Background subtraction is notably difficult and different methods can yield contrasting conclusions on the physical properties of loops \citep{Terzo2010}. 
In any case, by definition, the neighboring loops do not pulsate \citep[Figure 4 of][]{Froment2015, Auchere2016}. Therefore the pulsations would still be present after subtraction of a co-spatial background estimated from neighboring loops \citep[e.g.][]{Aschwanden2011}. In addition, since the automatically detected regions of excess Fourier power clearly take the shape of visible bundles of loops and of the corresponding extrapolated magnetic field lines~\citep[][]{Froment2016}, the detected pulsations can safely be attributed to these bundles of loops. The associated Fourier and wavelet power spectra present all the characteristics expected from random pulse trains (see \S~\ref{sec:pulse_trains_psds}): a broad hump centered on zero frequency, a primary peak of power a few tens of $\sigma$ above, and possibly the presence of higher-order harmonics.    


\subsection{Comparison With Simulated Pulse Trains\label{sec:comparison_simulations}}

\begin{figure*}
	\centering
		\includegraphics[width=\textwidth]{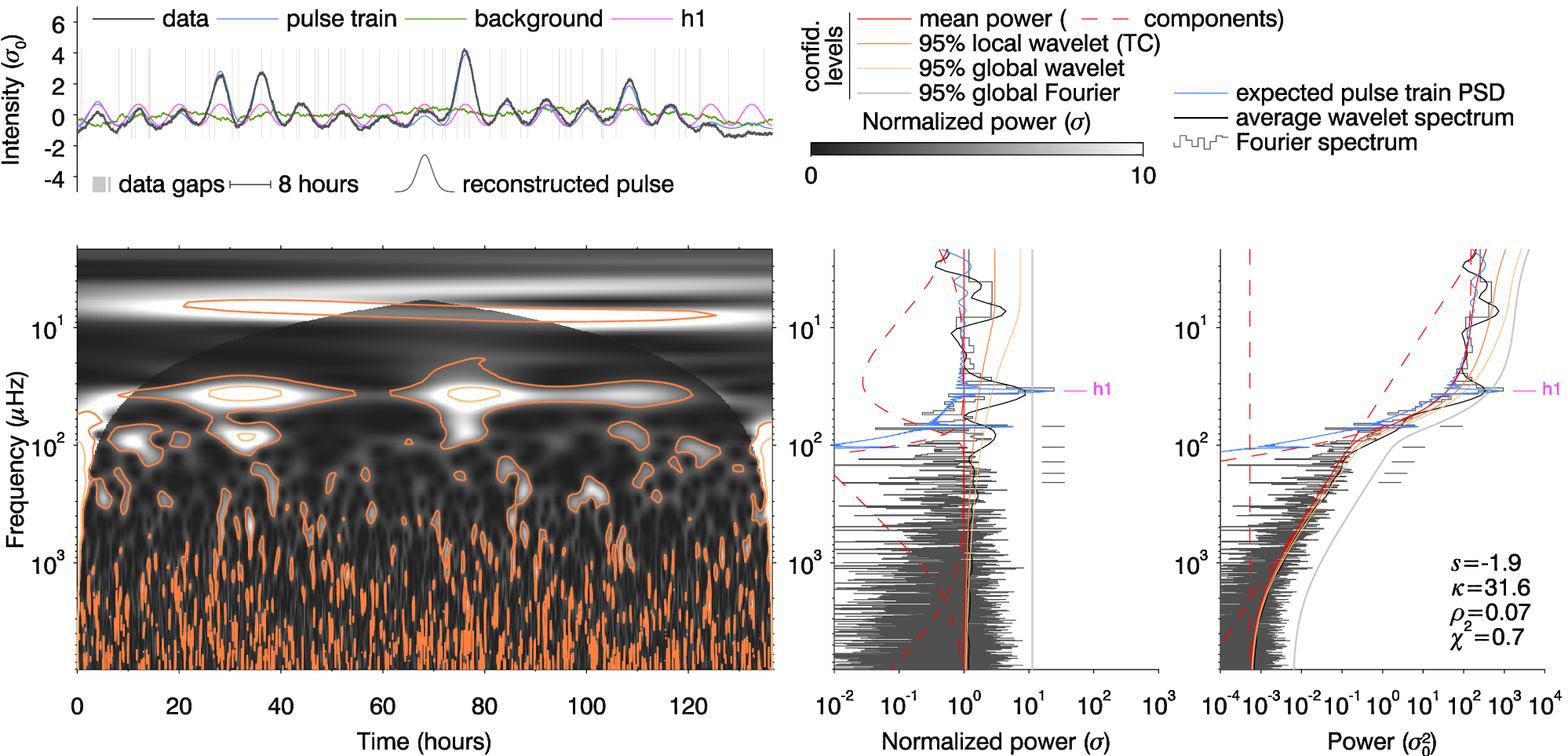}
		\caption{Fourier and wavelet analysis of simulated data based on a train of nearly Gaussian pulses of random amplitudes. This Figure is to be compared with Figure~\ref{fig:case_1_335}. See text for details.}
	\label{fig:sim_wavelet_gauss}
\end{figure*}

\begin{figure*}
	\centering
		\includegraphics[width=\textwidth]{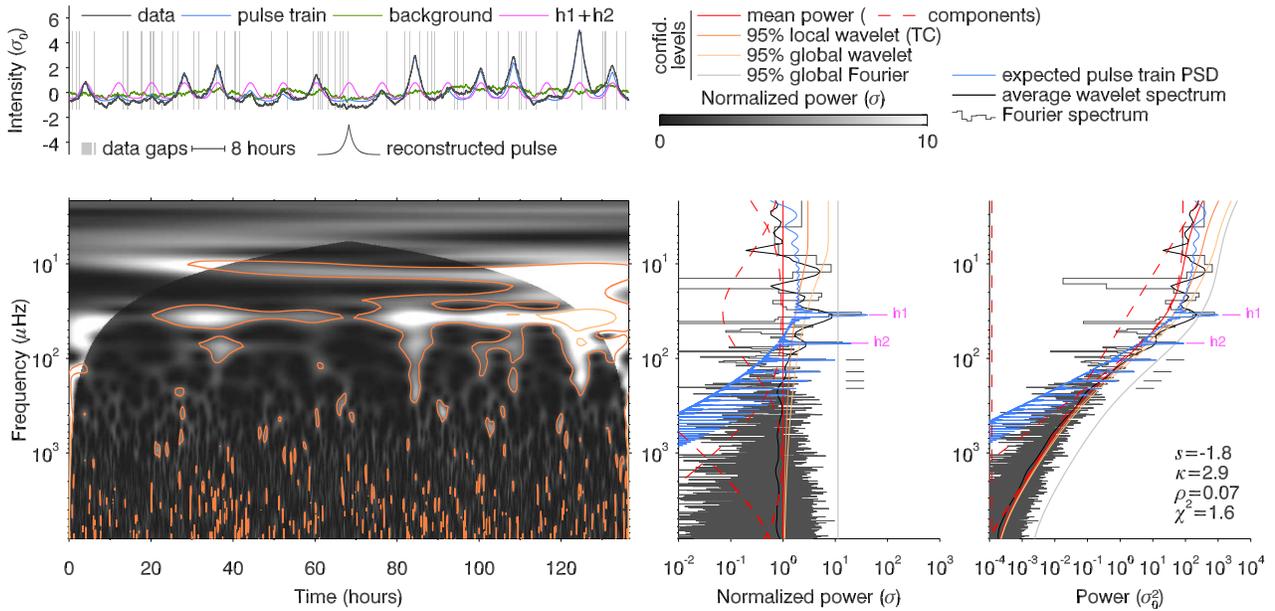}
		\caption{Same as Figure~\ref{fig:sim_wavelet_gauss} for a nearly double-exponential pulse train. To be compared with Figure~\ref{fig:case_3_171}.}
		\label{fig:sim_wavelet_expo}
\end{figure*}

In order to determine whether, and under what conditions, the fundamental and harmonic peaks expected in the PSDs of random pulse trains can be detected in real data, we simulated observations of the two pulse trains of Figure~\ref{fig:sim_pulse_train} by adding background emissions and photon noise, and we analyzed the resulting light curves using the exact same code as the real time series of \S~\ref{sec:aia_detection}. We set $T=8$~hr, a cadence $\delta t$ of 1 minute and a total of $N=8192$ data points, i.e. a total duration of 137~hr or 5.7 days. As for real data (see Equation~\ref{eq:kappa_model}), the PSDs of the simulated data have three components: the PSD of the pulse train, that of the background emission, and a constant produced by photon noise. 
The pulse trains and the background emissions were scaled so that the relative variances of the three components are similar to those in observed PSDs. The variance of the photon noise is equal to the mean of the signal, which was set to be comparable to that in the real AIA data (3~s exposures, $4\times 4$ binned images, summation over 231 heliographic pixels and 16~photons.s$^{-1}.\text{pixel}^{-1}$ at 33.5~nm, summation over 55~pixels and 750~photons.s$^{-1}.\text{pixel}^{-1}$ at 17.1 nm). The background emissions are random time series synthesized using the algorithm of \citet{Timmer1995} to have PSDs following power laws of exponent -2. The zero-mean backgrounds were scaled to have variances 672 and 128 times that of the photon noise at 17.1~nm and 33.5~nm respectively (higher signal-to-noise ratio at 17.1~nm than at 33.5~nm). The zero-mean pulse trains were normalized to both have variances ten times that of their respective background. Next we included photon noise by replacing the intensity in the total signal at each time step by a random deviate drawn from a Poisson distribution with that mean. Finally, we removed 30 randomly chosen data points to mimic data gaps.

The resulting light curves, normalized to their standard deviation $\sigma_0$, are shown in gray in the top left panels of Figures~\ref{fig:sim_wavelet_gauss} and~\ref{fig:sim_wavelet_expo} for, respectively, the nearly Gaussian ($\kappa=50$) and nearly double-exponential ($\kappa=5$) cases. The background emissions are shown in green and the pulse trains (identical to those of Figure~\ref{fig:sim_pulse_train}) in blue. These two figures are to be compared individually with Figures~\ref{fig:case_1_335} and~\ref{fig:case_3_171}:

\begin{enumerate}
\item In the bottom right panels, the Fourier (gray histograms) and global wavelet (black lines) spectra have identical shapes in simulated and real data: an overall power-law behavior flattening out at high frequencies with a hump between 10 and 100~$\mu$Hz. The latter is more pronounced in the AIA 33.5~nm and Gaussian pulse train spectra (Figures~\ref{fig:case_1_335} and~\ref{fig:sim_wavelet_gauss}). In simulated data, the hump matches the expected PSDs of the pulse trains\footnote{We use the following normalizations for the Fourier transform and the fast Fourier transform (FFT): $F(\nu)=\int_{-\infty}^\infty f(x)\exp\left(-2\ipi\nu x\right) \mathrm{d}x$, $\mathrm{FFT}(k)=\frac{1}{N}\sum_{n=0}^{N-1}x_n\exp\left(-2\ipi kn/N\right)$. The analytic PSDs are scaled by $(N\delta t)^2$ to match those computed by FFT.} (superimposed in blue and shown in the bottom panel of Figure~\ref{fig:sim_pulse_train}). 

\item A fundamental frequency (marked h1 in magenta) is detected in the Fourier spectra in all cases with comparable significance levels (10-30 times the local mean power). A second harmonic (marked h2) is also detected for the AIA 17.1~nm series and for the double-exponential train (Figures~\ref{fig:case_3_171} and~\ref{fig:sim_wavelet_expo}).

\item The model of Equation~\ref{eq:kappa_model} is a good fit to the mean power in all cases, as shown by the reduced $\chi^2$ values and by the flatness of the whitened spectra (middle panels). In the simulations, while the $s=-2$ slope of the power law of background emission and the width of the hump ($\rho=0.07$) are correctly recovered, the values of $\kappa$ differ significantly from the input. The reason is that the parameters of the kappa function are constrained only over a limited range of frequencies. Nonetheless, the fit correctly identifies the simulated pulses as nearly Gaussian ($\kappa=31.6$, Figure~\ref{fig:sim_wavelet_gauss}) and nearly double-exponential ($\kappa=2.9$), Figure~\ref{fig:sim_wavelet_expo}), as shown also by the elementary pulses reconstructed by inverse Fourier transformation of the kappa function component. 

\item From Equation~\ref{eq:kappa_psd} and Figure~\ref{fig:sim_pulse_train}, the more peaked the pulses, the more extended is the high-frequency wing of their PSD. It is thus easier to detect high order harmonics for peaked pulses than for rounded ones. The hump in Figures~\ref{fig:case_1_335} and~\ref{fig:sim_wavelet_gauss} drops too rapidly ($\kappa=29.3$ and $\kappa=31.6$) for the second harmonic to be detected. Conversely, in Figures~\ref{fig:case_3_171} and~\ref{fig:sim_wavelet_expo} , the extended wing ($\kappa=2.6$ and $\kappa=2.9$) remains above the power-law component for longer and the second harmonic is visible. 

\item The high-frequency wing of the kappa function is itself a power-law (Equation~\ref{eq:kappa_psd}), so it can be difficult to distinguish the hump from the background in the case of peaked pulses (Figures~\ref{fig:case_3_171} and~\ref{fig:sim_wavelet_expo}). 

\item Significant power is detected in all the wavelet spectra (bottom left panels), but intermittently. Indeed, the power in each pulse scales with the square of its amplitude, while the model of background power used to derive the confidence levels is constant with time.

\item The strongest peaks of power in the wavelet spectra of Figures~\ref{fig:sim_wavelet_gauss} and~\ref{fig:sim_wavelet_expo} present high-frequency extensions (most visible between 25 and 60~hr in the AIA 17.1~nm data). These correspond to the enhanced visibility of the high-frequency wing of the PSD of strong individual pulses with respect to the background power law. 
\end{enumerate}

This comparison shows that the fundamental characteristics of the Fourier and wavelet spectra of Figure~\ref{fig:case_1_335} (respectively Figure~\ref{fig:case_3_171}) can be explained by the presence of a nearly Gaussian (respectively nearly double-exponential) pulse train in the AIA 33.5~nm (respectively 17.1~nm) time series. 

\section{CONCLUSIONS\label{sec:conclusions}}

Numerical simulations of coronal loops indicate that periodic thermal non-equilibrium cycles are an unambiguous tracer of quasi-steady footpoint heating. TNE has been proposed as a viable explanation of the intensity pulsations that we recently detected in coronal loops~\citep{Froment2015, Froment2016}.  Since the boundary conditions relevant to TNE (loop geometry, heating rate, and localization, etc.) are likely to vary randomly over time, it is expected that each TNE cycle will be different from the preceding one, effectively producing periodic intensity pulses of random amplitudes. In this paper, we demonstrated that the PSDs of the time series reported by~\cite{Auchere2014} and~\cite{Froment2015} indeed exhibit the characteristic harmonics and continuum expected from random pulse trains. We thus explicitly use the terminology {\it periodic pulses}, as opposed to {\it oscillations}, which would incorrectly suggest that the observed periodicities correspond to vibrational modes. The theoretical PSD of pulse trains to which we compared our observations presupposed that the amplitudes are not correlated (see the Appendix). However, correlated amplitudes -- e.g. resulting from a remnant of the conditions of past cycles  -- would only modify the contrast between the harmonic peaks and the continuum of the PSD~\citep{Xiong2000}. In all cases, the harmonics are the signature of the periodicity of the pulses, the continuum is the signature of the randomness of their amplitudes, and the ratio between the two constrains the PDF of the latter.



The identification of random pulse trains in the data reinforces TNE as being the correct explanation for the slow pulsations observed in coronal loops. \citet{Auchere2014} estimated that half of the active regions in the year 2000 underwent a pulsation event. Considering that many events may have been missed by the automatic detection algorithm (because of, e.g., the high detection thresholds, data gaps, or the bias toward strictly periodic events inherent to working in the Fourier space), it is reasonable to think that the vast majority of active regions exhibit this type of behavior at least once in their lifetime. In addition, using one-dimensional hydrodynamic simulations with realistic loop geometries from photospheric magnetic field extrapolations, \citet{Froment2016} have shown that the region of parameter space for which TNE cycles develop is very limited, thus explaining why only some of the loops of an active region exhibit pulsations even if all were heated quasi-steadily at their footpoints. Since TNE is already the standard model of prominence formation and coronal rain, we now have a growing body of evidence that quasi-steady footpoint heating is more common in active regions than previously thought, even though the fundamental mechanism could still be anything from truly continuous wave dissipation to high-frequency nanoflares. As a final point, about half of the pulsation events reported by \citet{Auchere2014} were located in the quiet Sun, which tantalizingly hints that TNE may be at play in these regions too.

\begin{acknowledgements}
The authors acknowledge the use of the wavelet code by \cite{Torrence1998}. The authors acknowledge the use of of \sdoaia data. This work used data provided by the MEDOC data and operations centre (CNES/CNRS /Univ. Paris-Sud), http://medoc.ias.u-psud.fr/.
\end{acknowledgements}

\bibliographystyle{apj}
\bibliography{bibliography}

\appendix

\section{PSD OF RANDOM-AMPLITUDE PERIODIC PULSE TRAINS\label{sec:pulse_trains}}

A pulse train $f(t)$ formed by a succession at regular intervals~$T$ of $M$ copies of varying amplitudes $a_m$ of an elementary pulse $p(t)$ is given by
\begin{equation} 
f(t) = \sum_{m=0}^{M-1} a_m p(t - mT).
\label{eq:pulse_train}
\end{equation}


\noindent
There is particular interest in the situation where the amplitudes $a_m$ result from a stochastic process, in which case $f(t)$ is called a random pulse train.\footnote{ \cite{Lucht2013} introduced instead the term \emph{statistical pulse train} to describe an idealized pulse train whose statistical properties would match those of a sufficiently large number of realizations of the random amplitudes.} Indeed, the signals transmitted in communication systems are trains of symbols whose occurrence is practically random. The statistical properties of the Power Spectral Density (PSD) of random pulse trains have thus been studied extensively since at least the 1950s \citep[e.g.][]{Kaufman1955, Huggins1957, Barnard1964, Beutler1968} and are described in engineering textbooks \citep{Vincent1973, Xiong2000} because they condition the optimization of the usage of transmission bandwidth. Equation~\ref{eq:pulse_train} corresponds to the analog information encoding scheme called pulse-amplitude modulation in telecommunications.\footnote{Other encoding schemes correspond to other types of pulse trains for which the location or the width of the individual pulses is modulated \citep{Kaufman1955, Beutler1968}.} For completeness and for the convenience of the reader, we re-derive below the expression of the corresponding PSD as used in the main body of the present paper.

\noindent
If the elementary pulse $p(t)$ is square-integrable, then, as long as $M$ is finite, $f(t)$ is square-integrable as well and thus has finite total power. Using the property of linearity, we write the Fourier transform of $f(t)$ as the sum of the Fourier transforms of the individual pulses:
\begin{equation}
F(\nu)=\sum_{m=0}^{M-1}\int_{-\infty}^{+\infty}a_m p(t - mT)\ \text{e}^{-2\ipi\nu t}dt.
\label{eq:generic_pulse_train_fourier_transform}
\end{equation}

\noindent
Using the time-shifting theorem on Equation~\ref{eq:generic_pulse_train_fourier_transform}, we obtain the following expression for the PSD $\abs{F(\nu)}^2$ of $f(t)$
\begin{equation}
\abs{F(\nu)}^2=\abs{P(\nu)}^2\abs{\sum_{m=0}^{M-1} a_m \text{e}^{-2\ipi\nu mT}}^2,
\label{eq:generic_pulse_train_power}
\end{equation}

\noindent
where $P(\nu)$ is the Fourier transform of the elementary pulse $p(t)$ and $\abs{P(\nu)}^2$ is its PSD. We note that the summation in Equation~\ref{eq:generic_pulse_train_power} is the Fourier transform\footnote{This transform is the continuous transform of a discrete signal, sometimes called the discrete-time Fourier transform (DTFT), not to be confused with the discrete Fourier transform, which is obtained by evaluating the DTFT at discrete frequencies.} 
of the list of $a_m$. The PSD of $f(t)$ is thus the product of the PSD of the elementary pulse and the PSD of the discrete amplitudes.\footnote{Up to now we have in fact not made assumptions on the amplitudes. Therefore Equation~\ref{eq:generic_pulse_train_power} is valid for any decomposition of a function $f(t)$ on the basis of functions $p(t - mT)$.}

 
%

Since the amplitudes $a_m$ are random, the PSD varies randomly for different realizations of the random process. However, we can compute the {\it expected} PSD, i.e. the statistical average of the PSDs corresponding to an infinite number of these realizations. Expanding Equation~\ref{eq:generic_pulse_train_power} and using the commutativity property of finite sums, we have
\begin{eqnarray}
\Psi(\nu) & = & \mathrm{E}\left[\abs{F(\nu)}^2\right]\nonumber\\
 & = & \abs{P(\nu)}^2\mathrm{E}\left[\abs{\sum_{m=0}^{M-1} a_m  \text{e}^{-2\ipi\nu mT}}^2\right] \nonumber\\
 & = & \abs{P(\nu)}^2\mathrm{E}\left[\left(\sum_{m=0}^{M-1} a_m \text{e}^{-2\ipi\nu mT}\right)\left(\sum_{m=0}^{M-1} a_m \text{e}^{-2\ipi\nu mT}\right)^*\right] \nonumber\\
 & = & \abs{P(\nu)}^2\sum_{l=0}^{M-1}\sum_{m=0}^{M-1}\mathrm{E}\left[ a_l a_m\right] \text{e}^{-2\ipi\nu(l-m)T},
 \label{eq:expected_power_development}
\end{eqnarray}

\noindent
where $\mathrm{E}$ is the expected-value operator and $^*$ denotes the complex conjugate.\footnote{Note that $\mathrm{E}\left[ a_l a_m\right]$ is the autocorrelation of the amplitudes.} Assuming stationarity of the random process causing the amplitude variations\footnote{It is worth noting that the mean $\mu$ and the variance $\sigma^2$ are not the mean and the variance of the $M$ amplitudes $a_m$. Each $a_m$ is a random variable whose mean and variance could be estimated from a large number of realizations. For a stationary process, the mean and variance of the ensemble of $M$ amplitudes $a_m$ would tend to $\mu$ and $\sigma^2$, respectively, for an infinite pulse train.}, all of the $a_m$ have the same mean $\mu = \mathrm{E}\left[a_m\right]$ and the same variance $\sigma^2$. If we further add the assumption that the $a_m$ are independent\footnote{The case where the amplitudes are correlated is treated in, e.g., \citet{Huggins1957, Barnard1964, Xiong2000}. \cite{Barnard1964} details the specific case of Markov pulse trains, for which each amplitude depends only on that of the previous pulse.}, then we also have $\mathrm{E}\left[ a_l a_m\right]=\mathrm{E}\left[ a_l\right]\mathrm{E}\left[ a_m\right]$. Thus,

\begin{equation}
\mathrm{E}\left[a_l a_m\right] =
\begin{cases}
\mathrm{E}\left[a_l^2\right]=\sigma^2+\mu^2 & \text{if $l=m$}\\
\mathrm{E}\left[a_l a_m\right]=\mu^2 & \text{if $l\neq m$}
\end{cases}.
\end{equation}

\noindent
Substituting these expressions into Equation~\ref{eq:expected_power_development} by splitting the sums into the cases $l=m$ and $l\neq m$, we get
\begin{eqnarray}
\Psi(\nu) & = & \abs{P(\nu)}^2\left\lbrace M\sigma^2 + \mu^2\sum_{l=0}^{M-1}\sum_{m=0}^{M-1}\text{e}^{-2\ipi\nu(l-m)T}\right\rbrace \nonumber\\
& = & \abs{P(\nu)}^2\left\lbrace M\sigma^2 + \mu^2\left(\sum_{m=0}^{M-1} \text{e}^{-2\ipi\nu mT}\right)\left(\sum_{m=0}^{M-1} \text{e}^{-2\ipi\nu mT}\right)^*\right\rbrace.
\end{eqnarray}

\noindent
The two latter sums are geometric progressions and can thus be rewritten using the relation $\sum_{n=0}^{N-1}q^n=(q^N-1)/(q-1)$ \citep[relation 0.112, p. 1]{Gradshteyn1994}, from which it follows that
\begin{equation}
\Psi(\nu) = \abs{P(\nu)}^2\left\lbrace M\sigma^2 + \mu^2\left(\frac{\sin(\pi\nu TM)}{\sin(\pi\nu T)}\right)^2\right\rbrace,
\label{eq:random_pulse_train_power}
\end{equation}

\noindent
which is equivalent to Equation 35.17 of  \cite{Lucht2013}. The second term of Equation~\ref{eq:random_pulse_train_power} is periodic with period $1/T$. In the vicinity of zero, $\sin^2(\pi\nu TM)/\sin^2{\pi\nu T}\approx M^2\mathrm{sinc}^2(\pi\nu TM)$, which tends to $M\delta(\nu)/T$ when $M\to \infty$. The PSD given by~\ref{eq:random_pulse_train_power} diverges for an infinite number of pulses, but we can define $\Psi'(\nu) = \Psi(\nu)/M$ which is the PSD {\it per pulse} so that
\begin{equation}
\lim_{M\to\infty}\Psi'(\nu) = \abs{P(\nu)}^2\left\lbrace \sigma^2 + \frac{\mu^2}{T}\sum_{m=-\infty}^\infty\delta\left(\nu - \frac{m}{T}\right)\right\rbrace,
\label{eq:infinite_random_pulse_train_power}
\end{equation}
\noindent
which is identical to Equation A.17 of \cite{Xiong2000} \citep[see also Equations (2a), (2b) and Table III of ][]{Kaufman1955}. Under the hypotheses of stationarity and independence, the expected PSD of a random pulse train is thus the product of the PSD of the elementary pulse and the sum of two components: a constant and a periodically peaked function that tends to a Dirac comb when the number of pulses is large. 
Finally, if the amplitudes are constant then $\sigma^2=0$ and the second member of Equation~\ref{eq:infinite_random_pulse_train_power} reduces to the product of the PSD of the elementary pulse and a Dirac comb of period $1/T$, a result that could have been obtained directly from Equation~\ref{eq:pulse_train} using the convolution theorem.

%
%

\end{document}